\documentclass[doublecol,figures]{epl2}
\usepackage{amsmath}
\usepackage{amssymb}
\usepackage{amsfonts}
\usepackage{bm}
\usepackage{dsfont}
\usepackage{mciteplus}

\newcommand{\s}{\sum\limits}

\newcommand{\pa}{\partial}

\newcommand{\be}{\begin{equation}}
\newcommand{\e}{\end{equation}}
\newcommand{\beml}{\begin{subequations}}
\newcommand{\eml}{\end{subequations}}
\newcommand{\beq}{\begin{eqnarray}}
\newcommand{\eq}{\end{eqnarray}}
\newcommand{\ba}{\begin{array}}
\newcommand{\ea}{\end{array}}
\newcommand{\lt}{\left}
\newcommand{\rt}{\right}

\newcommand{\la}{\langle}
\newcommand{\ra}{\rangle}
\newcommand{\ep}{\varepsilon}

\newcommand{\bb}{\boldsymbol}

\newcommand{\h}{^\dagger}
\newcommand{\ph}{^{\phantom{\dagger}}}
\newcommand{\0}{^{\mbox{ }}}

\DeclareMathOperator{\tr}{Tr}
\DeclareMathOperator{\diag}{diag}
\DeclareMathOperator{\im}{Im}

\title{Ballistic charge transport in chiral-symmetric few-layer graphene}

\author{W.-R. Hannes\inst{1} 
\and M. Titov\inst{1,2} 
}
\shortauthor{W.-R. Hannes and M. Titov}
\institute{
	\inst{1} School of Engineering \& Physical Sciences, Heriot-Watt University, Edinburgh EH14 4AS, UK \\
	\inst{2} DFG Center for Functional Nanostructures, Universit\"at Karlsruhe, 76128 Karlsruhe, Germany
}


\abstract{
A transfer matrix approach to study ballistic charge transport 
in few-layer graphene with chiral-symmetric stacking configurations is developed. 
We demonstrate that the chiral symmetry justifies a non-Abelian gauge transformation
at the spectral degeneracy point (zero energy). This transformation proves 
the equivalence of zero-energy transport properties of the multilayer to those 
of the system of uncoupled monolayers. Similar transformation can be applied 
in order to gauge away an arbitrary magnetic field, weak strain, and hopping disorder 
in the bulk of the sample. Finally, we calculate the full-counting statistics 
at arbitrary energy for different stacking configurations. The predicted 
gate-voltage dependence of conductance and noise can be measured in 
clean multilayer samples with generic metallic leads. }

\begin{document}
\maketitle

\section{Introduction}
\hspace*{-2mm}Discovery of graphene \cite{Novoselov04} has revived an interest 
to a wide range of phenomena that are explicitly or implicitly linked 
to the symmetry of the band structure \cite{Kane05,*Kane05a,Moore07}.
Graphene and topological metals realised 
in BiSb compounds \cite{Hsieh08} are very recent examples where 
exotic band-structure properties are responsible for unusual 
quantum effects in the charge transport. The well-known one is 
the integer quantum Hall effect that takes on remarkably distinct forms in
mono- and bi-layer graphene \cite{Novoselov05,Zhang05,Novoselov06}.
The continuing development of device fabrication technology 
promises interesting applications of these effects in microelectronics, 
chemical sensing, and quantum information. 

At the same time there is a growing interest in new functionalities 
of few-layer graphene that is inspired by the proposals to control valley 
polarisation \cite{Rycerz07,Abergel09},  band gaps \cite{McCann06}, 
and magnetic exchange \cite{Semenov08} with the help of gate electrodes. 
The experimental evidence supporting these theoretical proposals 
is currently building up \cite{Oostinga08,Han09}. 
Few-layer graphene is also a prime candidate for the observation 
of excitonic condensation caused by an attractive 
interlayer Coulomb interaction \cite{Min08a,Kharitonov08}.

\begin{figure}[t]
\onefigure[width=0.43\textwidth]{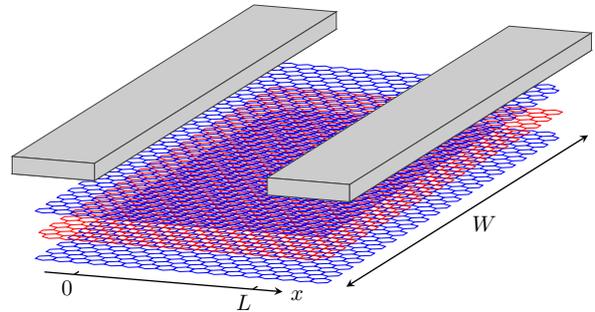}
\caption{Schematic view of a rectangular trilayer graphene sample contacted by metallic leads.}
\label{fig:setup}
\vspace*{-3mm}
\end{figure}

In this letter we develop a unified approach to transport properties 
of chiral-symmetric multilayer setups shown in fig.~\ref{fig:setup}. 
Our main result is the universality of the full-counting statistics 
for the charge transport at the spectral degeneracy point ($\ep=0$).
We demonstrate that this universality is the consequence 
of the chiral symmetry of the tight-binding model of graphene multilayer. 
Therefore, it extends to arbitrary perturbations of the model, 
which preserve the chiral symmetry. Such perturbations include: 
weak crystal strains, arbitrary magnetic field, and interlayer hopping disorder.

\begin{figure}[t!]
\onefigure[width=0.43\textwidth]{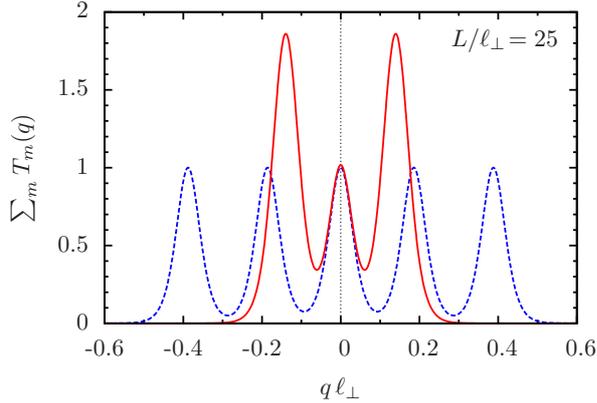}
\caption{The sum of transmission probabilities, $\sum_{m=1}^M T_m(q)$, at $\ep=0$ 
as a function of the transversal momentum $q$ in the case of few layer graphene ($M=5$)  
with Bernal (solid line) and rhombohedral (dashed line) stacking and vanishing magnetic field.
}
\label{fig:Tq}
\end{figure}

The universal charge transport is provided by the evanescent modes 
that originate at the metal-graphene boundaries.  We find that 
the universal two-terminal conductivity of the setup is given 
by $\sigma = 4 M e^2/\pi h$, where $M$ is the number of layers, 
$e$ is the electron charge, and $h$ is the Planck constant. 
The Fano factor at zero energy takes on the universal value $F=1/3$. 
These results are insensitive to magnetic field, weak strain, and hopping disorder,
which can be gauged away from the full-counting statistics 
by means of a non-Abelian gauge transformation. Individual 
transmission eigenvalues are, however, less universal and
strongly depend on the stacking configuration (see fig.~\ref{fig:Tq}).

\begin{figure}[t!]
\onefigure[width=0.43\textwidth]{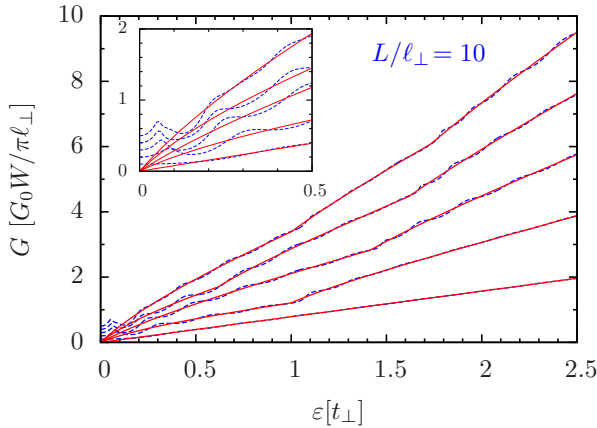}
\caption{Quantum-mechanical (dashed lines) and averaged conductance 
of AB-stacked few-layer graphene as a function of the Fermi-energy. 
The layer number is one through five from bottom to top. The inset is 
a zoom into the low-energy region.}
\label{fig:G}
\end{figure}

\begin{figure}[t!]
\onefigure[width=0.43\textwidth]{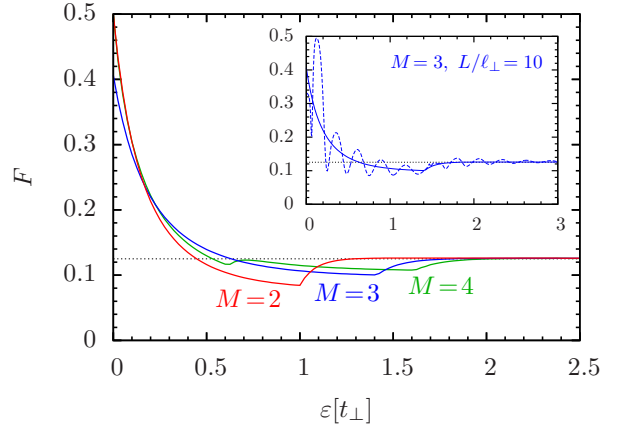}
\caption{Averaged Fano factor in transport through AB-stacked few-layer graphene. 
The horizontal line corresponds to $F=1/8$. The inset shows a comparison 
with the exact result (dashed curve) found from eqs.~(\ref{pichard},\ref{full},\ref{GF}).
}
\label{fig:F}
\end{figure}

\begin{figure}[t!]
\onefigure[width=0.43\textwidth]{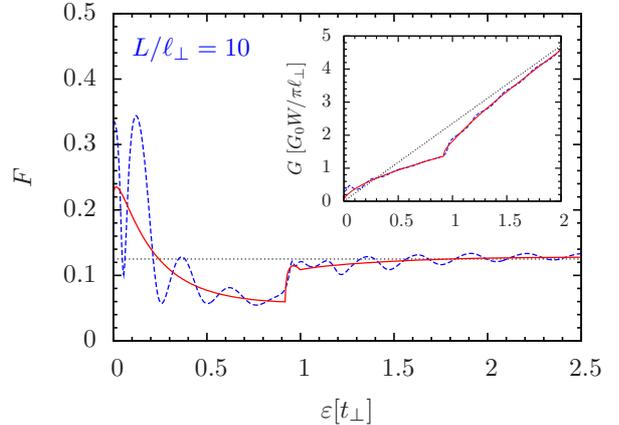}
\caption{Exact (dashed line) and averaged (solid line) Fano factor 
for ABC-stacked trilayer graphene. The inset shows the transmission spectrum. 
}
\label{fig:Fano-G_ABC}
\end{figure}

In the second part of the letter we adopt the scattering approach to 
calculate the full-counting statistics of the charge transport away
from the degeneracy point for $\ep \gg \hbar v/L$, where $L$ 
is the distance between the metallic leads and $v$ is the Fermi velocity.
This transport regime is dominated by the propagating modes that 
give rise to the sample-dependent Fabri-Perot oscillations 
in the conductance and noise of the ballistic setup. The gate-voltage 
dependence of these quantities is further analyzed by averaging 
over the Fabri-Perot oscillations. Our results for regular stacks are presented 
in figs.~\ref{fig:G}-\ref{fig:Fano-G_ABC}.
For any multilayer above the highest band threshold 
we find the averaged conductance $G= M e^2 W |\ep|/h$
and seemingly universal value of the Fano factor, $F=1/8$. 

\section{Model}

The chiral symmetry is preserved in the nearest-neighbour tight-binding model
of few-layer graphene for the regular stacking types, Bernal (AB$\cdots$) 
and rhombohedral (ABC$\cdots$), and for all irregular ones 
excluding sequences of the type AA.

For these models the Fermi surface becomes point-like 
at the spectrum degeneracy point ($\ep=0$). At this energy 
no propagating states exist, hence the conductivity of the idealised material
is expected to vanish. This is, however, not true in the setup of a finite size 
shown in fig.~\ref{fig:setup} due to the presence of zero-energy evanescent modes,
which contribute to the transport in a peculiar way 
\cite{Tworzydlo06,Katsnelson06,Snyman07,Titov07,Schuessler09,Titov09,Titov10}.
The theory predicts the conductance of a ballistic monolayer setup $G=\sigma W/L$, 
where the two-terminal conductivity, $\sigma$, approaches the universal value $4 e^2/\pi h$ 
at zero energy, while the Fano takes on the value $F=1/3$ as in a diffusive system. 
Such pseudo-diffusive transport regime has been indeed observed 
in experiments \cite{Miao07,Danneau08} with submicron monolayer flakes.

Here we extend and generalize these results for few-layer chiral-symmetric graphene.
The model Hamiltonian is given by the sum of four terms
$H=H_0+H_A+H_S+H_\perp$, where $H_0$ describes the system of $M$ 
uncoupled monolayers, $H_A$ and $H_S$ account for magnetic field
and weak internal strain, correspondingly, and $H_\perp$ takes into account
the coupling between the layers.

In the effective mass approximation the Hamiltonian of isolated monolayers,
\be
H_0= 
\mathds 1_M \otimes (-\mathrm{i}\hbar v\,{\bb \sigma}\cdot{\bb \nabla})
\otimes\mathds 1_2^{(s)}\otimes \mathds 1_2^{(v)}, 
\e
consists of $M$ copies of the familiar two-dimensional Dirac equation
with the velocity $v\approx 10^{6} \text{m/s}$ and the vector of Pauli matrices 
$\bb{\sigma}=(\sigma_x,\sigma_y)$. The notations $\mathds 1_2^{(s)}$ and $\mathds 1_2^{(v)}$ 
stand for the unit matrices in spin and valley space, respectively. 
The external magnetic field is described by the term $H_A=(ev/c)\, 
\bb{\sigma}\cdot \hat{\bb{A}}\otimes \mathds 1_2^{(s)}\otimes \mathds 1_2^{(v)}$,
where $\hat{\bb{A}}(\bb{r})={\rm diag}(\bb{A}_1,\bb{A}_2,\dots ,\bb{A}_M)$ with $\bb{A}_m$ 
is the vector potential in the $m$-th layer. 
The weak crystal strain (e.g. induced by ripples) can be taken into account 
by a similar term $H_S=\bb{\sigma}\cdot \hat{\bb{S}}\otimes \mathds 1_2^{(s)}\otimes \tau_z$ 
\cite{Pereira09} with a strain field, $\bb{S}(\bb{r})={\rm diag}(\bb{S}_1,\bb{S}_2,\dots ,\bb{S}_M)$, 
where $\tau_z$ stands for the Pauli matrix in the valley space.
To simplify the notations we omit outer products and let $\hbar v=ev/c=1$ 
in those expressions where it cannot cause confusion.

The interlayer coupling is described by taking into account the nearest-neighbour hopping 
both for intralayer and interlayer processes \cite{Koshino07,Min08b},
\be
H_\perp= \frac{t_\perp}{2} 
\lt(\ba{ccccc} 
0 & s\ph_{1} & 0 & 0 & \vdots\\
s\h_{1} & 0 & s\ph_{2} & 0 & \vdots\\
0 & s\h_{2} & 0 & s\ph_{3} & \vdots\\ 
0 & 0 & s\h_{3} & 0 & \vdots \\
\hdots & \hdots & \hdots & \hdots & \ddots
\ea\rt),
\e
with $t_\perp=$ 0.3-0.4\,eV and $s_n =\sigma^{\pm} = \sigma_x \pm \mathrm{i} \sigma_y$,
where the choice of sign depends on the stacking order. The energetically most favourable configuration 
is the Bernal stacking (AB) characterised by $s_{2m} =\sigma_+$,  $s_{2m+1} =\sigma_-$.
The less favourable rhombohedral configuration (ABC)
corresponds to $s_{m}=\sigma_+$ for $m=1,2,\dots M-1$. 
The model is readily generalised to account for spatial
variations of the hopping parameter.

The central property of our model is the chiral symmetry
\be
\label{chiral}
\sigma_z H \sigma_z=-H,
\e
which holds for any spatial dependence of the vector potential $\bb{A}$, 
strain field $\bb{S}$, and the hopping parameter $t_\perp$. 

It follows from the chiral symmetry (\ref{chiral}) that the spectrum determined 
by the "Dirac" equation, $H\Psi=\ep \Psi$, is symmetric with respect to $\ep=0$.
Analytical results are available for regular multilayers 
with $H_A=H_S=0$. In particular, the spectrum of the ABC-stacked multilayer
is gapless and contains two branches touching at $\ep=0$ \cite{Min08b},
\be
\label{ABC}
\ep_{\bb{k}}=\pm t_\perp (|\bb{k}|\ell_\perp)^M
+ t_\perp\, {\cal O}\lt((|\bb{k}|\ell_\perp)^{M+1}\rt),
\e
where $\ell_\perp\equiv\hbar v/t_\perp =$ 1.6-2.2\,nm and $\bb{k}$ is the wave-vector.
The spectrum of AB-stacked multilayer is given by \cite{Guinea06}
\be
\label{ABA}
\ep_{m,\bb{k}}=t_\perp c_m \pm t_\perp \sqrt{|\bb{k}|^2 \ell_\perp^2 +c_m^2}, 
\quad c_m=\cos\frac{\pi m}{M+1},
\e
where $m=1,2,\dots,M$. 
The result (\ref{ABA})
is due to the exact mapping \cite{Koshino07} onto a system
of $M/2$ bilayers (for even $M$) or onto a system of a single monolayer 
and $(M-1)/2$ bilayers (for odd $M$).
 
For both models (\ref{ABC}) and (\ref{ABA}) the Fermi surface 
at $\ep=0$ becomes point-like. We shall see that this property extends to a general model 
with arbitrary spatially dependent terms $H_A$, $H_S$ and $H_\perp$. 

We define the matrix $\hat{\Omega}(\bb{r})$, which yields 
the equation $-\bb{\sigma \nabla}\hat{\Omega} = H_A+H_S+H_\perp$
and introduce the local gauge transformation \cite{Aharonov79}
\be
\label{gauge}
\Psi(\bb{r}) = \mathrm{e}^{\mathrm{i}\hat{\Omega}(\bb{r})}\Phi(\bb{r}).
\e

Using the chiral symmetry (\ref{chiral}) one can show that the transformation (\ref{gauge})
relates the zero-energy eigenstate, $\Psi_0$, of the full Hamiltonian, $H\Psi_0=0$, 
to the eigenstate, $\Phi_0$, of the bare system, $H_0 \Phi_0=0$. 
Since the spectrum of $H_0$ is gapless at $\ep=0$ and the Fermi surface 
is degenerate at $\ep=0$, the same properties apply to the full model, $H$.

In the following sections we explore the consequences for the transport properties 
of graphene multilayer.

\section{Transfer matrix approach}

We apply the scattering approach to the transport in graphene multilayers 
for the two-terminal setup depicted in fig.~\ref{fig:setup} 
disregarding interaction effects. In particular, we calculate 
the full-counting statistics for the charge transport at zero frequency 
and temperature, which is determined 
by the cumulant generating function \cite{Levitov93} 
\be
\label{full}
{\cal F}(\chi) =  \ln \det \lt[1-\hat{t} \hat{t}\h+\mathrm{e}^\chi \,\hat{t} \hat{t}\h \rt],
\e
where $\hat{t}$ is the matrix of transmission amplitudes for the scattering states in the leads
at a given energy $\ep$. The conductance and the Fano factor are given by, respectively,
\be
\label{GF}
G=G_0\lim_{\chi\to 0} \frac{\pa {\cal F}}{\pa \chi}, \quad\ 
F=(G_0/G)\lim_{\chi\to 0} \frac{\pa^2 {\cal F}}{\pa \chi^2}, 
\e
where $G_0=4e^2/h$ is the conductance quantum (the factor 
of 4 takes into account the spin and the valley degeneracy). 

Following earlier works \cite{Tworzydlo06,Snyman07,Titov07} we apply sharp boundary conditions 
at the metal-graphene interfaces at $x=0$ and $x=L$. We also restrict ourselves to the limit
$W\gg L$, where $W$ is the width and $L$ is the length of the rectangular graphene sample.
Without loss of generality we apply periodic boundary conditions in $y$ direction,
(a particular form of the boundary conditions is irrelevant for $W \gg L$), 
hence the momentum quantization in the leads, $q_n=2\pi n/W$, 
where $q$ stands for the projection of the momentum on $y$ axis. 

The transport properties are, then, characterised by the evolution matrix, ${\cal T}$,
which relates the wave-function Fourier components in the left and the right lead  
$\Psi_q(L)=\sum_{q'}{\cal T}_{qq'}\Psi_{q'}(0)$. In the case of rigid boundary conditions 
the relation between the evolution matrix and the transfer matrix, ${\cal M}$, 
takes the simple form \cite{Titov07}, ${\cal M}={\cal L}\h {\cal T} {\cal L}$ 
with ${\cal L}=(\sigma_z+\sigma_x)/\sqrt{2}$. 
Using the relation
\be
\label{pichard}
\frac{{\cal M}{\cal M}\h}{(1+{\cal M}{\cal M}\h)^2} =
\frac{1}{4}\lt(\!\!\! \ba{cc} {\hat{t} \hat{t}\h}&0\\0&\hat{t}^{\prime\dagger}\hat{t}^\prime\ea\!\!\!\rt),
\e 
we rewrite the cumulant generating function (\ref{full}) as
\be
\label{full2}
{\cal F}=\frac{1}{2}\ln \det \lt[\lt(1-{\cal T}{\cal T}\h\rt)^2+4\mathrm{e}^\chi \,{\cal T}{\cal T}\h \rt].
\e
In the next section we use this form of the generating function to analyse the transport properties
of chiral graphene multilayers at $\ep=0$. 

\section{Transport at zero energy} 
The transformation (\ref{gauge}) enables us to calculate
the charge transport at $\ep=0$ for a general model. Let us start, however,
from a simple case of ballistic multilayer in the absence of magnetic 
and strain fields, $H_A=H_S=0$. In this case, the gauge transformation (\ref{gauge}) 
reduces to $\Psi=\mathrm{e}^{\mathrm{i} \hat{\Sigma} x} \Phi$ 
with $\hat{\Sigma} = -\sigma_x  H_\perp$. Using this transformation 
we obtain the evolution matrix as
\be
\label{E0transfer}
{\cal T}=\mathrm{e}^{q L\sigma_z}\mathrm{e}^{\mathrm{i} \hat{\Sigma} L},\quad\ 
{\cal T}{\cal T}\h = \lt(\!\!\! \ba{cc} \mathrm{e}^{2 q L} \hat{P} & 0 \\ 0 & \mathrm{e}^{-2 q L} \hat{P}^{-1} \ea\!\!\! \rt),
\e
with $\diag(\hat{P},\hat{P}^{-1}) = \mathrm{e}^{\mathrm{i} \hat{\Sigma} L}\mathrm{e}^{-\mathrm{i} \hat{\Sigma}^\dagger L}$.
We used that the matrices $\sigma_z$ and $\hat{\Sigma}$ commute. 
The eigenvalues of the matrix $\hat{P}$ are 
parameterised by $\exp(-2\kappa_mL)$. 
Since the eigenvalues coincide with those
of $\hat{P}^{-1}$, they appear in 
pairs with $\kappa_{m,m'}=\pm | \kappa_m |$. 
(The unpaired eigenvalue for odd $M$ corresponds to $\kappa_m = 0$.)
From equations (\ref{E0transfer}), (\ref{pichard}) one finds individual transmission 
probabilities  (which are  the eigenvalues of $\hat{t}\hat{t}\h$) as
\be
\label{T0}
T_{m}(q)=\lt[\cosh^{2}(q-\kappa_m)L\rt]^{-1}, \quad\ m=1,2,\dots M.
\e
The values of $\kappa_m$ play the role of momentum shifts, which
are irrelevant in the limit $W\gg L$, since the quantization of $q$ is dense.
In this limit the zero-energy generating function is obtained from equation (\ref{full2}) as
\be
\label{resF}
{\cal F}(\chi)=-M (W/\pi L) \lt[\arccos (\exp{\chi/2})\rt]^2. 
\e
This form of the generating function coincides with that of a diffusive system 
\cite{Dorokhov82} despite the ballistic nature of charge transport 
in our model. 
The direct consequence of equation (\ref{resF}) is the universal form 
of the conductivity, $G$, and the Fano factor, $F$, at zero energy
\be
\label{zeroE}
G=4 e^2 M W/\pi h L, \qquad\ F=1/3.
\e
The full-counting statistics (\ref{resF}) and the results (\ref{zeroE}) hold irrespective of the 
stacking order between the layers and are independent of the stacking 
specific momentum shifts $\kappa_m$. 
These results rely on the validity of 
the sharp boundary conditions with
${\rm max}\,\kappa_m \ll  2\pi/\lambda_F$, where $\lambda_F$ stands 
for the Fermi wave length in the lead. 

The momentum dependence of the transmission probabilities is
less universal which is illustrated in fig.~\ref{fig:Tq}.
For a multilayer with AB stacking we find 
\be
\label{qABA}
\kappa_m=L^{-1}\ln\lt(c_m L/\ell_\perp+\sqrt{1+c_m^2 (L/\ell_\perp)^2}\rt),
\e
where the coefficients $c_m$ are defined in eq.~(\ref{ABA}) and $m=1,2,\dots, M$.
This result is consistent with the mapping of the multilayer onto independent bilayers 
\cite{Koshino07}. For $M=2$, the result of eq.~(\ref{qABA}) has been obtained by Snyman 
and Beenakker \cite{Snyman07}. 

For the ABC stacking configuration we find the asymptotic expressions 
\be
\label{qABC}
\kappa_m=L^{-1}\ln\lt(\frac{(M-m)!}{(m-1)!}(L/\ell_\perp)^{2m-M-1}\rt),
\e
in the limit $\ell_\perp\ll L$. The transmission resonances shown 
in fig.~\ref{fig:Tq} are much better separated for the ABC multilayer 
than for the multilayer with the Bernal stacking. 
The momentum shifts $\kappa_m$ in both cases depend logarithmically 
on the ratio $L/\ell_\perp$, so that the condition ${\rm max}\,\kappa_m \ll  2\pi/\lambda_F$, 
is hard to violate. The validity of eq.~(\ref{T0}) is restricted to $|\ep|\ll \ep_0$, 
where $\ep_{0}=t_\perp /(2c_1) (\pi\ell_\perp/L)^2$ for AB and
$\ep_{0}\approx t_\perp (\pi\ell_\perp/L)^M$ for ABC stacking.
For ballistic graphene ribbons with $W\lesssim L$, the
full counting statistics is sensitive to the shifts $\kappa_m$ due 
to the transversal momentum quantization.

Remarkably, the results (\ref{resF},\ref{zeroE}) remain valid 
even in the presence of arbitrary magnetic and strain fields.
To justify this statement it is convenient to consider 
the evolution operator ${\cal T}$ in the real space representation, 
such that $\Psi(L,y)={\cal T}\Psi(0,y)$. Using the transformation (\ref{gauge}) 
we find
\be
\label{magnetic}
{\cal T} = \mathrm{e}^{\mathrm{i}\hat{\Omega}(L,y)}
\mathrm{e}^{-\mathrm{i}\sigma_z L\pa_y} \mathrm{e}^{-\mathrm{i}\hat{\Omega}(0,y)},
\e 
where the magnetic and strain fields and inter-layer coupling are entering solely
by means of the matrix phase $\hat{\Omega}$ taken at the graphene-metal boundaries. 
The gauge of the vector potential can be chosen such that the phase at the boundary is $y$-independent. 
Hence the matrix exponents in eq.~(\ref{magnetic}) commute and the evolution operator 
in the channel space takes the form of eq.~(\ref{E0transfer})
with $\Sigma=(\hat{\Omega}(L)-\hat{\Omega}(0))/L$.

\section{Gate-voltage dependence}

In the vicinity of the Dirac point, for $\ep\ll \hbar v/L$, the transport is entirely 
due to the evanescent modes, which are responsible for the 
pseudo-diffusive form of the full counting statistics (\ref{resF}) at $\ep=0$. 
Away from the Dirac point, the transport is less universal and depends 
on a number of details. In this section we restrict ourselves 
to ballistic models with $H_A=H_S=0$. 

We calculate the transmission probabilities $T_m(q)$ and perform 
the numerical integration over the transverse momentum $q$ 
(assuming $2\pi/\lambda_F\gg |\ep|$) 
to find the conductance and the Fano factor in the limit $W\gg L$. 
In the case of AB stacking the probabilities $T_m(q)$ can be found analytically
and the resulting energy dependence of the conductance is shown 
in fig.~\ref{fig:G} with dashed lines. The Fano factor for the trilayer with AB and ABC 
stacking is depicted with the dashed lines in fig.~\ref{fig:F} (inset) 
and fig.~\ref{fig:Fano-G_ABC}, correspondingly. 
At $\ep=0$ the figures confirm the results of eqs.~(\ref{resF},\ref{zeroE}).

For energies exceeding the ballistic Thouless energy, $\hbar v/L$, 
the transport is dominated by propagating modes, which give rise to
the sample-specific Fabry-P\'{e}rot oscillations in conductance and noise. 
In order to get experimentally relevant results we perform 
the averaging over these oscillations treating the propagating phases 
as random quantities, which are uniformly distributed in the interval 
$(0,2\pi)$ \cite{Buttiker88,Blanter00}. This type of averaging corresponds 
to a quasiclassical approximation that respects the conservation 
of the transversal momentum, $q$, in the sample. 

We introduce the individual scattering matrices 
for the left (L) and right (R) sample-lead interfaces
\be
\label{Smatrices}
\hat S_\mathrm{L/R} = \lt(\ba{cc} \hat\rho_\mathrm{L/R} 
& \hat\tau'_\mathrm{L/R} \\ \hat\tau_\mathrm{L/R} & \hat\rho'_\mathrm{L/R} \ea\rt),
\e
which relate the scattering state amplitudes in the leads with 
those in the sample. We assume that the number of open channels in the sample
equals $M_0 \leq M$ for a given energy, $\ep$, and transversal momentum, $q$. 
In the planar geometry of fig.~\ref{fig:setup} the $S$-matrices 
in eq.~(\ref{Smatrices}) are readily calculated by matching 
the scattering states in the corresponding lead and in the sample.
 
As the result the total transmission matrix from the left to the right lead can be written as
\be 
\label{t-decomp}
\hat t = \hat \tau_\mathrm{R}  
\lt(\mathds 1 - \hat\tau_0\ \hat\rho_\mathrm{L}'\ \hat\tau_0\ 
\hat\rho_\mathrm{R} \rt)^{-1}\ \hat\tau_0\ \hat\tau_\mathrm{L} ,
\e
where $\hat{\tau}_0 = \mathrm{diag}(\mathrm{e}^{\mathrm{i} \phi_1},
\dots ,\mathrm{e}^{\mathrm{i} \phi_{M_0}})$ 
is parameterised by the propagating phases $\phi_m=k_m L$ accumulated 
in the free propagation inside the sample. The number of propagating channels, $M_0$, 
is determined from the requirement $\im(k_m) =0$, where the longitudinal momentum 
$k_m$ is found from the dispersion relation $\ep=\ep(k,q)$
for a given energy $\ep$ and transversal momentum $q$. 

An equivalent way to calculate the averaged conductance is formulated 
in terms of the classical transmission and reflection probabilities, 
$\mathrm{T}_{\mathrm{R(L)},nm} = |\tau_{\mathrm{R(L)},nm}|^2$,
and $\mathrm{R}_{\mathrm{R(L)},nm} = |\rho_{\mathrm{R(L)},nm}|^2$,
for the transport through the sample-graphene interfaces.
The classical probabilities to pass through an entire sample
are, then, organised in the following matrix 
$\mathrm{\hat{T}}=\mathrm{\hat{T}_R} 
(\mathds{1} - \mathrm{\hat{R}'_L}
\mathrm{\hat{R}_R})^{-1} 
\mathrm{\hat{T}_L}$,
for each value of $q$. The quantity, $\mathrm{\hat{T}}$,
ignores the phase-coherence in the assumption that the transversal 
momentum is conserved inside the sample. The averaged conductance, 
$\overline{G}=G_0 \sum_q\sum_{nm} \mathrm{T}_{nm}$ coincides with 
$\overline{G}=G_0 \sum_q \la \tr \hat{t}\hat{t}\h\ra$, where the brackets stay
for the averaging over the propagating phases $\phi_m$ 
in Eq.~(\ref{t-decomp}).

A special case of a single propagating channel, $M_0=1$, per transversal momentum, $q$,
is naturally realised in a monolayer and multilayers with a single ungapped band 
for energies below the lowest band threshold. The example of the latter is the ABC-stacked 
multilayer. For equivalent sample-lead junctions, the cumulant generating function (\ref{full}) 
can be expressed using eq.~(\ref{t-decomp}) as
\be
 {\cal F}(\chi) = \sum_q \ln \frac{\bar{T}^2 \mathrm{e}^\chi+4 (1-\bar{T}) \sin^2\phi}
 {\bar{T}^2+4 (1-\bar{T}) \sin^2\phi},
\e
where $\bar{T}=\hat{\tau} \hat{\tau}\h$ is the $q$-dependent transmission 
probability of the sample-lead interface. The averaging over the propagating phase, $\phi$, 
can be performed analytically with the result
\be
\label{Faveraged}
\overline{\cal F} (\chi) = 2 \sum_q 
\ln \lt[ \bar{T} \mathrm{e}^{\chi/2}+\sqrt{4(1-\bar{T})+\bar{T}^2\mathrm{e}^{\chi}}\rt].
\e
With the help of eq.~(\ref{GF}) we obtain for the averaged conductance and the Fano factor 
\be
\label{singlemode}
\overline{G}=G_0 \s_{q} \frac{\bar{T}}{2-\bar{T}},
\quad\ \overline{F}=\frac{G_0}{G}\s_{q}\frac{2\bar{T}(1-\bar{T})}{(2-\bar{T})^3}.
\e

Let us digest eqs.~(\ref{Faveraged},\ref{singlemode}) 
in the case of monolayer graphene. If the leads 
in the setup of fig.~(\ref{fig:setup}) are modelled by
highly doped graphene one finds the interface transmission 
probability as $\bar{T}(q)=1-\tan^2\theta/2$, 
where $q=|\ep|\sin\theta$ and $\theta\in (-\pi/2,\pi/2)$ 
is the angle of incidence. For $W\gg L$ we replace the summation 
in eqs.~(\ref{singlemode}) with the integration over $q$ in the interval $(-\ep,\ep)$
and reproduce the asymptotic results
\cite{Schuessler09} $\overline{G}=G_0 W |\ep|/4$ and $\overline{F}=1/8$.

Similarly, for the bilayer we find from eqs.~(\ref{Faveraged},\ref{singlemode})  
the averaged conductance below the band threshold, 
\begin{align} \label{Gbi}
\overline{G} &= G_0 W \lt( \frac{|\ep|}{4} + 
t_\perp^2 \frac{\gamma-\eta}{8|\ep|} \rt),
\quad\ |\ep| < t_\perp , 
\end{align}
where $\gamma = 1+2|\ep|/t_\perp$ and $\eta=\sqrt{1+\gamma-\gamma^{-1}}$. 
The result for the Fano factor in the bilayer setup takes the form
\begin{align} \label{Fbi}
\overline{F}&=
\frac{G_0 W}{\overline{G}} 
\lt( \frac{|\ep|}{32} + 
t_\perp^4 \frac{4 c_1+c_2/(\gamma^3\eta)}{512 |\ep|^3} \rt),
\quad\ |\ep| < t_\perp, 
\end{align}
where $c_1=3-7\gamma(1+\gamma)-\gamma^3$ and 
$c_2= 3+\gamma(\gamma+2)(16\gamma^4+12\gamma^3-17\gamma^2+10\gamma-6)$.

If several propagating channels per the value of $q$ open up in the sample, 
the averaging procedure is complicated and has to be carried out numerically. 
Still, for AB-stacked multilayers the analysis is simplified 
by using Koshino-Ando mapping \cite{Koshino07} to an effective bilayer-monolayer system.
For energies below the first band threshold, the conductance 
and noise can be constructed from the available analytical results for mono- 
and bi-layer graphene using the effective bilayer coupling constants, 
$t^{\mathrm{eff}}_{\perp,m}= 2 c_m t_\perp$, where $m=1,2,\dots,\mathrm{Int}[M/2]$. 
For higher energies, one finds at most two propagating channels per $q$ so that the  
numerical implementation of the averaging procedure is straightforward.
The band thresholds are seen as kinks in the 
energy dependence (transmission spectra) of conductance and noise 
in figs.~\ref{fig:G} and~\ref{fig:F}. 
At $\ep=0$ we formally find $\overline{G}=0$ and (for AB-stacked multilayer) 
$\overline{F}=1/2$ for even $M$ and $\overline{F}=(6 M - 5)/(12 M - 4)$
for odd number of layers, $M$. These results ignore the contribution of
evanescent modes.

The interlayer coupling becomes irrelevant for the transport
at energies far above all band thresholds. Universal asymptotic results, 
$G=M G_0 W |\ep|/4$ and $F=1/8$, are obtained in this limit  
for any combination of AB and ABC stacking. This universality
is due to the linear dispersion, $\ep\approx c_m t_\perp \pm \hbar v |\bb{k}|$,
of all spectral branches of our effective few-layer model at high energies.

One can modify the decomposition (\ref{t-decomp}) to account  
for evanescent modes, which correspond to imaginary values 
of the propagating phases, $\phi_m=k_mL$, and use 
an appropriate analytic continuation of the matrices $\hat{S}_\mathrm{L/R}$, 
which become non-unitary. We employ this approach to plot the exact conductance and 
Fano factor for ABC-stacked graphene in fig.~\ref{fig:Fano-G_ABC}.

\section{Conclusions}
A class of Hamiltonians describing few-layer graphene with 
spatially dependent strain and magnetic field obeys a chiral symmetry. 
This symmetry is responsible for the universal pseudo-diffusive charge transport 
via evanescent modes. The universality is understood on the basis of
a non-Abelian gauge transformation at zero energy. For finite energies
the full counting statistics of few-layer ballistic samples 
is calculated using the scattering approach. 
 
\acknowledgments
We thank S.~Gattenl\"{o}hner for valuable discussions.
WRH acknowledges the financial support from the Scottish Universities Physics Alliance. 

\bibliography{references}

\end{document}